\documentclass[11pt]{article}

\usepackage[utf8]{inputenc}
\usepackage[T1]{fontenc}
\usepackage[margin=1in]{geometry}
\usepackage{graphicx}
\usepackage{amsmath, amssymb}
\usepackage{hyperref}
\usepackage{natbib}
\usepackage{setspace}
\usepackage{caption}
\usepackage{booktabs}
\usepackage{float} 
\usepackage{subcaption} 
\usepackage{cleveref}
\usepackage{authblk}

\title{CosinorAge: Unified Python and Web Platform for Biological Age Estimation from Wearable- and Smartwatch-Based Activity Rhythms}

\date{\today}

\author[1,2]{Jinjoo Shim\thanks{Corresponding author: jinjooshim@hsph.harvard.edu}}
\author[2]{Jacob Hunecke}
\author[2,3]{Elgar Fleisch}
\author[2]{Filipe Barata}

\affil[1]{Department of Biostatistics, Harvard University, Cambridge, MA, USA}
\affil[2]{Centre for Digital Health Interventions, ETH Zurich, Zurich, Switzerland}
\affil[3]{Centre for Digital Health Interventions, University of St.Gallen, St.Gallen, Switzerland}

\begin{document}
\maketitle  
\section{Summary}

Every day, millions of people worldwide track their steps, sleep, and activity rhythms with smartwatches and fitness trackers. These continuously collected data streams present a remarkable opportunity to transform routine self-tracking into meaningful health insights that enable individuals to understand—and potentially influence—their biological aging. Yet most tools for analyzing wearable data remain fragmented, proprietary, and inaccessible, creating a major barrier between this vast reservoir of personal health information and its translation into actionable insights on aging.\\

\texttt{CosinorAge} is an open-source framework that estimates biological age from wearable-derived circadian, physical activity, and sleep metrics. It addresses the lack of unified, reproducible pipelines for jointly analyzing rest–activity rhythmicity, physical activity, and sleep, and linking them to health outcomes. The Python package provides an end-to-end workflow from raw data ingestion and preprocessing to feature computation and biological age estimation, supporting multiple input sources across wearables and smartwatch. It also makes available trained model parameters (open weights) derived from large-scale population datasets such as UK Biobank, enabling reproducibility, transparency, and generalizability across studies. Its companion web-based \texttt{CosinorAge} Calculator enables non-technical users to access identical analytical capabilities through an intuitive interface. By combining transparent, reproducible analysis with broad accessibility, \texttt{CosinorAge} advances scalable, personalized health monitoring and bridges digital health technologies with biological aging research.

\section{Statement of Need}

Circadian rhythms play a critical role in maintaining key regulatory systems, including metabolic, immune, and endocrine pathways, and tightly govern rest–activity cycles encompassing sleep and physical activity, both essential to healthy aging. Disruptions in these daily rhythms, such as reduced amplitude, irregular activity timing, low activity levels, or poor sleep regularity, have been consistently linked to increased risk of chronic diseases, mortality, systemic inflammation, and accelerated biological aging \cite{shim2024circadian, shim2025wrist}. Given these established links, there is an urgent need for continuous high-resolution monitoring of daily rest-activity patterns to characterize individualized rhythmicity profiles, detect early deviations that signal elevated risk, and guide timely targeted interventions to optimize health span and slow biological aging.\\ 

Wearable devices and smartwatches enable a scalable, non-invasive, and cost-efficient method for digital biomarkers of circadian rhythms, physical activity, and sleep at both individual and population levels. However, existing analytic tools to analyze wearable data typically focus on isolated metric extraction or rely on proprietary algorithms, which limits transparency, reproducibility, and their ability to inform downstream health outcomes, such as biological age. To address this gap, we developed \texttt{CosinorAge} \cite{shim2024circadian}, a digital biomarker framework that estimates biological age and healthspan from circadian rest-activity rhythms using wearables.\\ 

Currently, there is no open-source analytic pipeline that offers a unified, end-to-end workflow for processing raw wearable data from data reading and pre-processing to feature computation and biological age estimation. Although circadian rhythm, physical activity, and sleep are physiologically interdependent and should be analyzed jointly, existing packages such as pyActigraphy \cite{hammad2021pyactigraphy}, actipy \cite{actipy}, CosinorPy \cite{movskon2020cosinorpy}, and scikit-digital-health \cite{adamowicz2022scikit} are limited to analyzing specific domains. While the GGIR R package \cite{ggir} provides extensive analysis of accelerometer data, it is only available in R and does not provide functionality to further linking metrics to health outcomes. Moreover, these tools are designed primarily for technical audiences, with no accessible interface for non-experts to upload and analyze their own data. This lack of accessibility limits the translational potential of scientific insights and leaves individuals dependent on opaque, proprietary manufacturer apps with limited interpretability.\\

\begin{figure}[H]
    \centering
    \includegraphics[width=\linewidth]{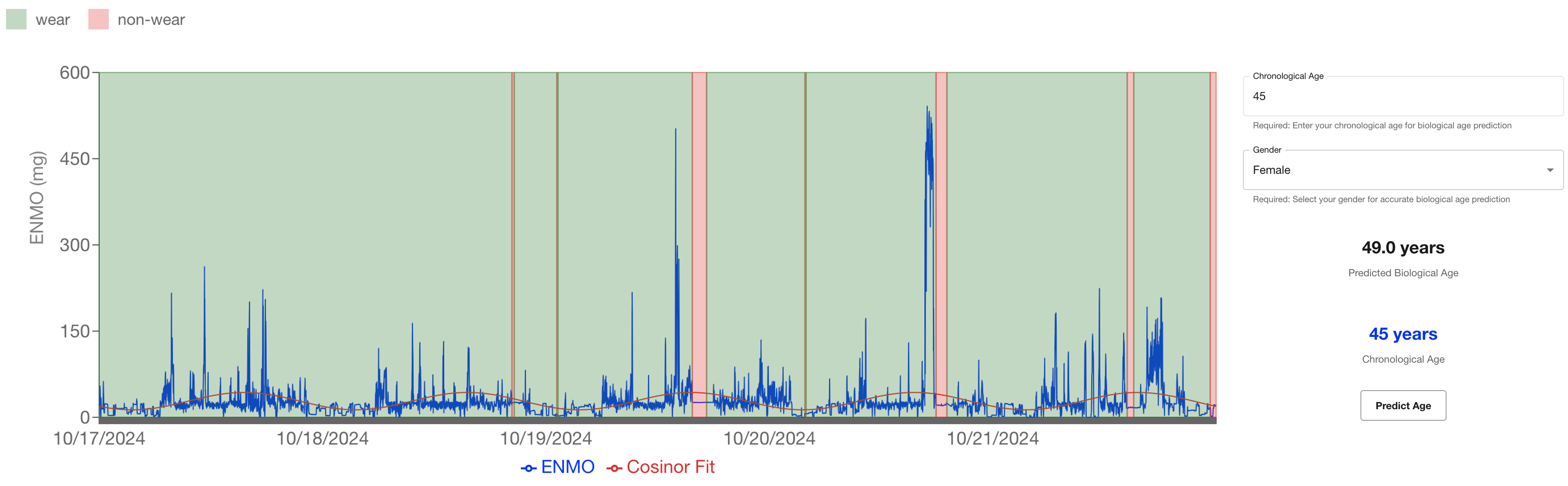}
    \caption{Minute-level activity data collected using a Samsung Galaxy smartwatch from a 45-year-old female over 7 days was analyzed using \texttt{CosinorAge} Python package. The blue lines display ENMO (Euclidean Norm Minus One) activity intensity (unit: mg), while the red curve indicates the cosinor model fit. Green and red shading mark wear and non-wear periods, respectively. Based on the recorded activity pattern, the predicted biological age is 49.0 years.}
    \label{fig:Fig1}
\end{figure}

In response, we developed the \textbf{CosinorAge Python Package} and \textbf{CosinorAge Calculator}. The Python Package provides a fully integrated workflow that processes raw accelerometer data, extracts features related to circadian rhythms, physical activity, and sleep, and estimates biological age, \texttt{CosinorAge}, as shown in Figure~\ref{fig:Fig1}. Importantly, \texttt{CosinorAge} represents a completely digital \textit{second-order clock}, derived from mortality risk rather than chronological age, thereby capturing healthspan-relevant biological processes with greater precision \cite{shim2024circadian}. The workflow supports input from a range of research-grade and consumer wearable devices, including research-grade actigraphy, large-scale population resources such as UK Biobank and NHANES, as well as consumer-centric smartwatches like the Samsung Galaxy Watch, thereby demonstrating broad compatibility across hardware platforms. Recent work has further validated the comparability between research-grade actigraphy and consumer-centric smartwatches in assessing circadian rhythms, underscoring the potential of \texttt{CosinorAge} to integrate data across different device types and study contexts \cite{wu2025comp}.\\

What makes our contribution particularly valuable is its scalability and accessibility: wearables are already widespread, cost-efficient, and continuously collecting high-resolution behavioral data. By leveraging this infrastructure, \textbf{CosinorAge Calculator} allows anyone--even without technical expertise--to estimate their biological age from their own wearable data. This democratization of health analytics opens a path to personalized monitoring of aging and resilience, bridging the gap between population-level research and personal health insights. Beyond individual users, these tools provide a common framework that unifies and extends methodologies across circadian rhythm, aging, and digital health. In doing so, these tools promote cross-disciplinary research and will be of broad interest to the scientific community. Together, these tools enhance reproducibility, promote accessibility, and expand the potential for continuous, data-driven health monitoring in aging research for both the scientific community and the general public.\\

A key aspect of \texttt{CosinorAge} is the release of the model coefficients derived from large-scale population datasets such as UK Biobank. Making these open weights available allows biological age estimation without retraining and facilitates comparability across new cohorts. This design provides a transparent alternative to proprietary algorithms and increases the translational utility of the package in both research and applied contexts.\\

The \textbf{CosinorAge Python Package} and \textbf{CosinorAge Calculator} are applicable to studies where wearable-derived behavioral rhythms are central to the research question. They can be used to quantify circadian, activity, and sleep patterns in terms of strength, timing, and stability, for example in studies examining whether rhythm disruptions are associated with aging trajectories or adverse health outcomes. When the research question concerns biological age, the package provides estimates from wearable data that can be incorporated into analyses of rhythm characteristics, aging processes, or intervention effects. The tools can process data from single individuals as well as multi-participant cohorts, providing both personal-level metrics and cohort-level summaries of behavioral rhythms and biological age that can be linked to broader health and aging research.\\

\section{How it Works}

\subsection{CosinorAge Python Package. } 
\begin{figure}[H]
    \centering
    \includegraphics[width=\linewidth]{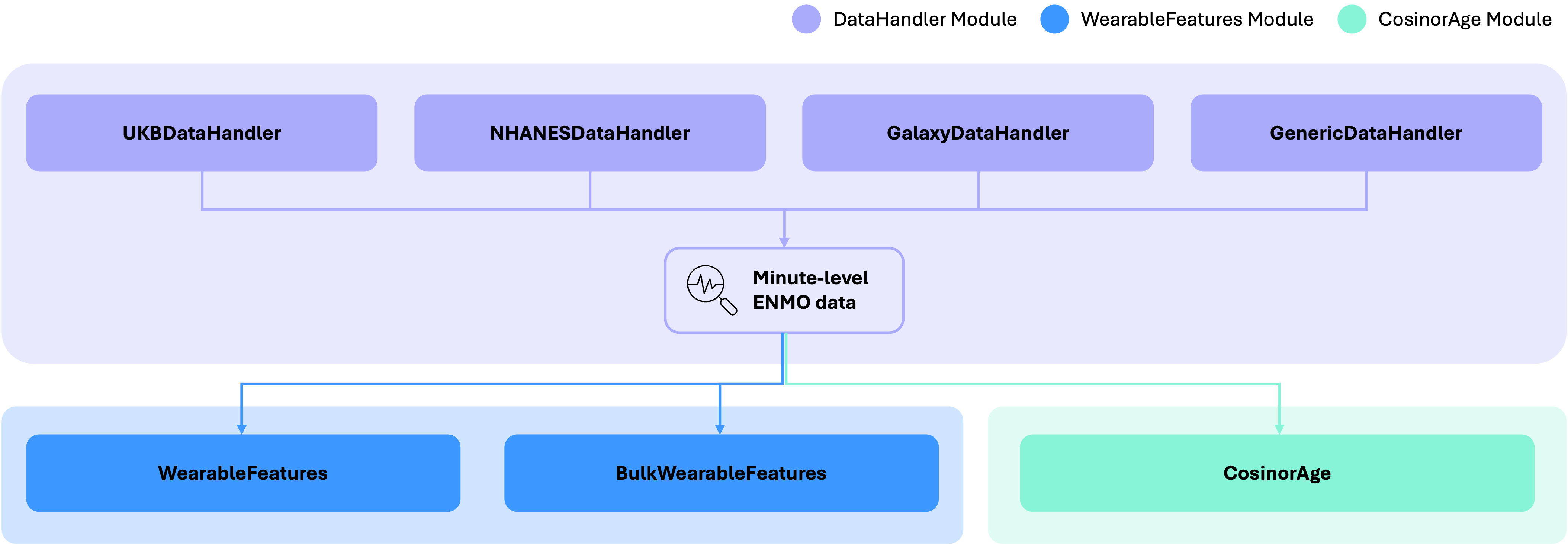}
    \caption{Package Scheme}
    \label{fig:Fig2}
\end{figure}

The \textbf{CosinorAge Python package} is structured into three core modules, each representing a key stage in the pipeline for analyzing accelerometer data and predicting biological age, \texttt{CosinorAge}. Its modular architecture allows components to be used independently or integrated into a streamlined workflow. Figure~\ref{fig:Fig2} illustrates the modular design and high-level data flow between components.

\paragraph{DataHandler Module}  
The package provides a total of four DataHandler subclasses to support accelerometer data from multiple sources including UK Biobank (UKB), NHANES, Samsung Galaxy Smartwatches (Galaxy), and Bring-Your-Own-Data (BYOD). UKBDataHandler, NHANESDataHandler, and GalaxyDataHandler perform source-specific filtering, preprocessing, and scaling to produce standardized, minute-level ENMO time series. Detailed data pre-processing for each DataHandler can be found on GitHub. For greater flexibility, a GenericDataHandler is also provided, allowing users to process any compatible CSV file formatted according to a defined specification through BYOD approach. The resulting ENMO data can then be passed to the feature extraction and modeling modules for downstream analysis.

\paragraph{WearableFeatures Module}  
The WearableFeatures module includes two classes: WearableFeatures and BulkWearableFeatures. Designed for individual-level analysis, the WearableFeatures class computes a comprehensive set of metrics from minute-level ENMO data, covering physical activity, sleep behavior, and both parametric and non-parametric circadian rhythm features. For cohort-level studies, the BulkWearableFeatures class supports batch processing of multiple individuals, enabling users to analyze feature distributions and explore inter-feature correlations across the population. The list of features computed from this module is summarized below:

\begin{table}[ht]
\small
\centering
\begin{tabular}{|p{4.7cm}|p{9.2cm}|}
\hline
\textbf{Domain} & \textbf{Metrics} \\
\hline
Circadian Rhythm Analysis & MESOR, cosinor amplitude, acrophase, M10, L5, interdaily stability (IS), intradaily variability (IV), relative amplitude (RA) \\
\hline
Physical Activity Analysis & Light physical activity (LPA), Moderate physical activity (MPA),vigorous physical activity (VPA), sedentary duration \\
\hline
Sleep Analysis & Total sleep time (TST), wake after sleep onset (WASO), percent time asleep (PTA), number of waking bouts (NWB), sleep onset latency (SOL) \\
\hline
\end{tabular}
\caption{Summary of extracted metrics by analysis domain}
\end{table}

\paragraph{CosinorAge Module}
The \texttt{CosinorAge} module represents the final stage of the pipeline and contains a single class responsible for predicting the \texttt{CosinorAge} biomarker. It takes minute-level ENMO data as input and applies a pre-trained proportional hazards model to estimate biological age \cite{shim2024circadian}. The model supports three sets of coefficients - unisex, female-specific, and male-specific. If available, sex can be included as an optional input to improve prediction accuracy. The underlying model coefficients were estimated from large-scale cohorts such as UK Biobank and are openly available. This open-weight design enables researchers to apply the same model across diverse datasets with clear and accessible parameters, thereby facilitating reproducibility and offering a transparent alternative to proprietary algorithms.

\subsection{CosinorAge Calculator: Web User Interface. } To enhance the accessibility of the \textbf{CosinorAge Python package}, we developed a web-based interface that allows researchers and users to analyze their own data without requiring any installation or programming expertise (www.cosinorage.app). Users can simply upload their data, which is processed by the \textbf{CosinorAge package} in the backend. Results are presented in a clear, report-style format that includes visualizations to aid interpretation. A multi-user mode is also available, enabling researchers to upload and analyze data from multiple individuals simultaneously, allowing for the exploration of feature distributions and correlations across cohorts.

The Web Interface is organized into several sections:
\begin{itemize}
    \item The “Home” tab provides an overview of the \texttt{CosinorAge} framework, its purpose, key features, and demo video.
    \item The “Documentation” tab offers comprehensive API and interface documentation.
    \item The “Calculator” tab hosts the core analysis workspace with interactive tools for uploading wearable data, running activity rhythm analyses and biological age estimation, and viewing results in real time.
    \item The “About” tab presents information about the research group and contributing members.
\end{itemize}

\begin{figure}[H]
    \centering
    \begin{minipage}[t]{0.49\linewidth}
        \centering
        \includegraphics[width=\linewidth]{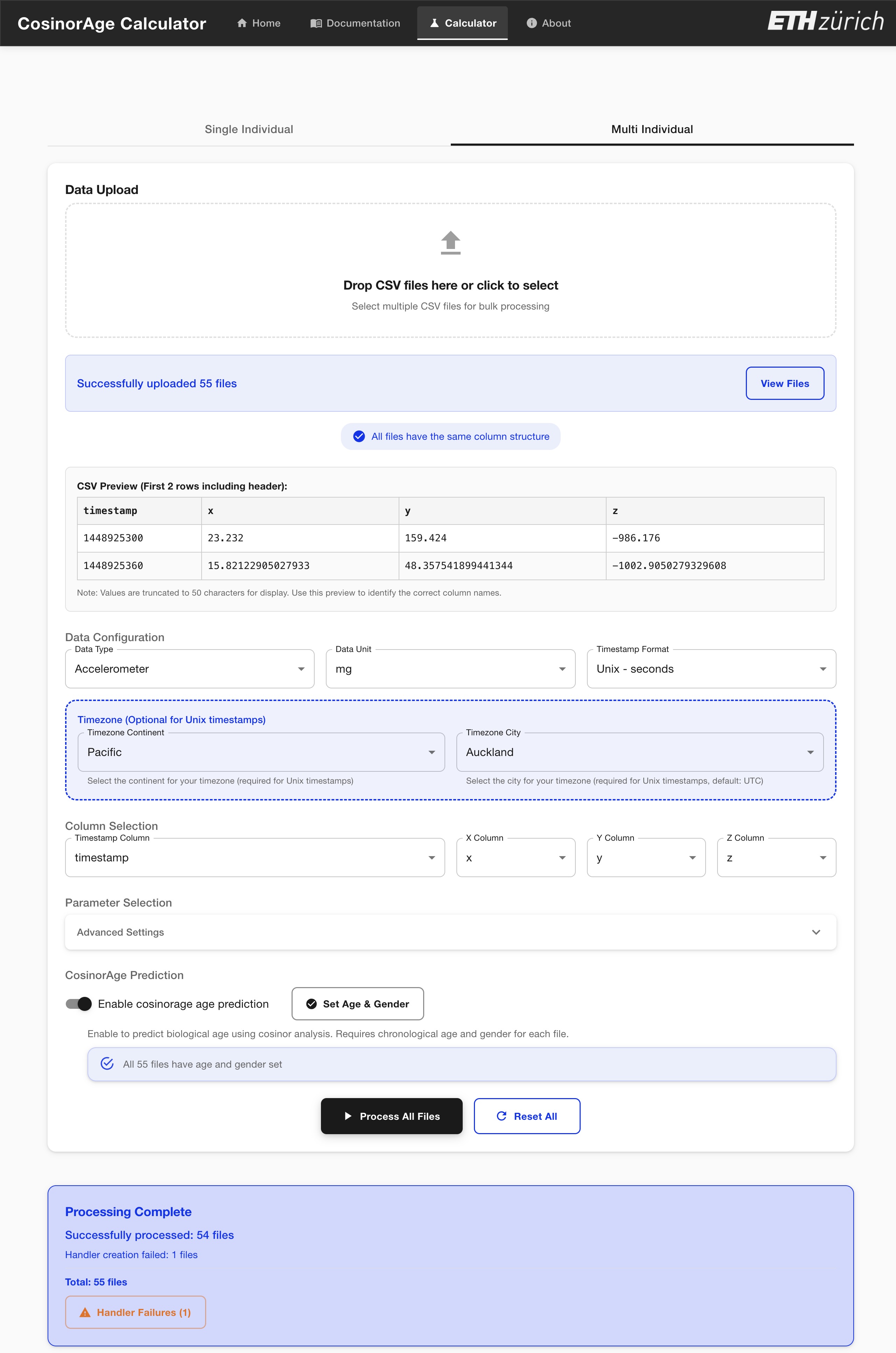}
        \caption{Data upload interface}
        \label{fig:Fig3}
    \end{minipage}
    \hfill
    \begin{minipage}[t]{0.49\linewidth}
        \centering
        \includegraphics[width=\linewidth]{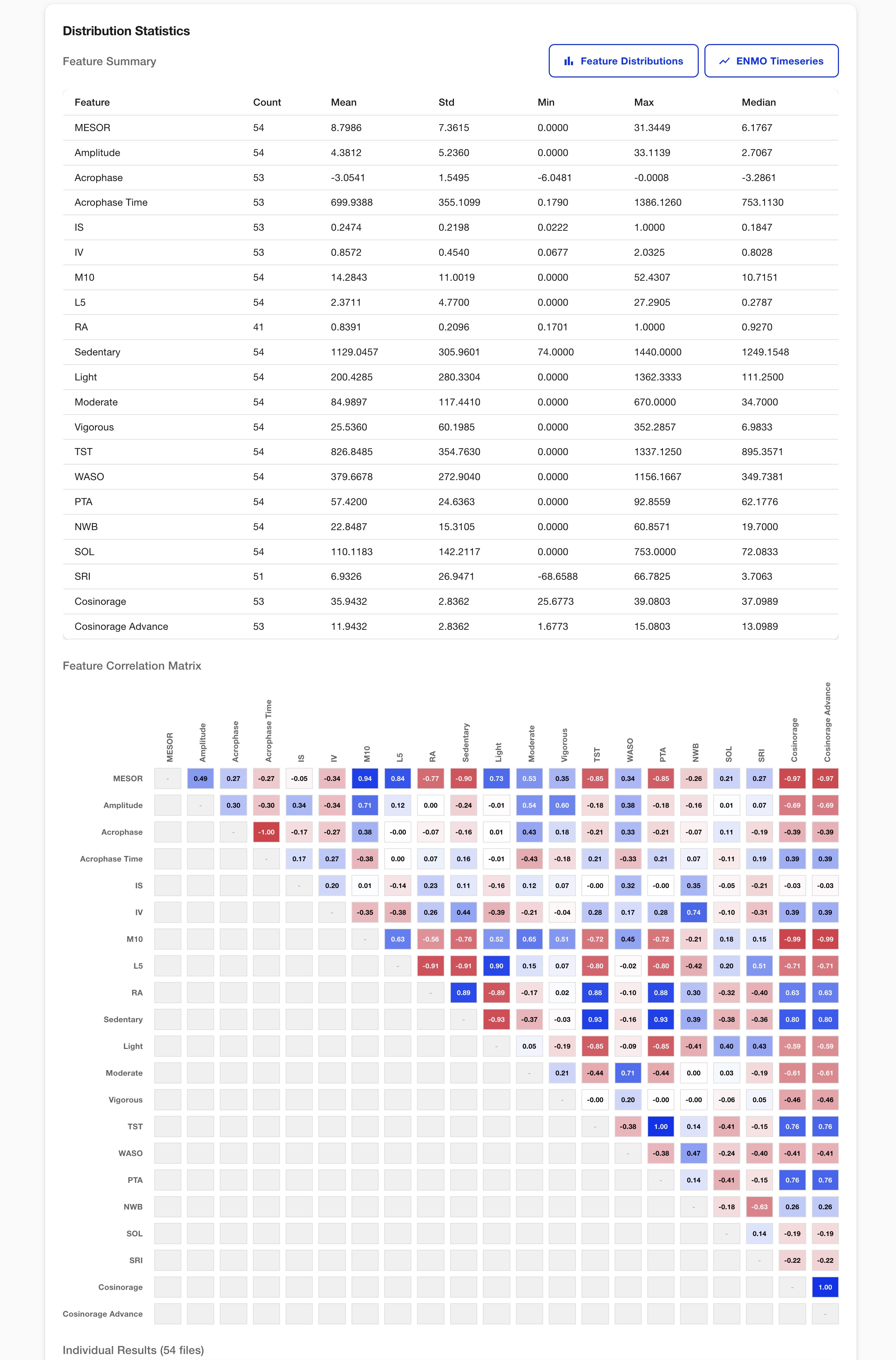}
        \caption{Summary dashboard}
        \label{fig:Fig4}
    \end{minipage}
\end{figure}

The "Calculator" tab offers a user-friendly interface, as illustrated in Figures~\ref{fig:Fig3} and \ref{fig:Fig4}. \textbf{CosinorAge Calculator} supports BYOD via batch CSV uploads from either single or multiple individuals, with automatic file structure preview for validation (subject to file size limits). Users can configure device type, timestamp format, time zone, and select parameters for analysis. When multi-individual mode is selected, the summary dashboard presents descriptive statistics for all extracted features, a feature correlation matrix, and visual summaries of each metric at the population level.

\bibliographystyle{abbrv}
\bibliography{references}

\end{document}